\documentclass[12pt]{article}

\usepackage{epsfig}

\usepackage{amssymb}
\usepackage{amsfonts}

\usepackage{color}
 
\def\be{\begin{equation}}
\def\ee{\end{equation}}
\def\ba{\begin{array}{c}}
\def\ea{\end{array}}

\newcommand{\bbr}{\br\!\br}
\newcommand{\kkt}{\kt\!\kt}

\newcommand{\kt}{\rangle}
\newcommand{\br}{\langle}

\begin{document}

\begin{center}

{\Large \bf
Intrinsic exceptional point -- a challenge in quantum theory
 }

\vspace{1.8cm}

  {\bf Miloslav Znojil}

\vspace{0.2cm}

\vspace{0.2cm}

\vspace{1mm} The Czech Academy of Sciences, Nuclear Physics
Institute,

Hlavn\'{\i} 130, 25068 \v{R}e\v{z}, Czech Republic,

\vspace{0.2cm}

 and

\vspace{0.2cm}

Institute of System Science, Durban University of Technology,

P. O.
Box 1334, Durban 4000, South Africa,

\vspace{0.2cm}

 and

\vspace{0.2cm}

Department of Physics, Faculty of Science, University of Hradec
Kr\'{a}lov\'{e},

Rokitansk\'{e}ho 62, 50003 Hradec Kr\'{a}lov\'{e},
 Czech Republic

%
%

\vspace{0.2cm}

 {e-mail: znojil@ujf.cas.cz}

\vspace{0.3cm}


\end{center}


\section*{Abstract}

In spite of its unbroken ${\cal PT}-$symmetry,
the popular imaginary cubic oscillator Hamiltonian
$H^{(IC)}=p^2+{\rm i}x^3$ does not satisfy
all of the necessary postulates of quantum mechanics.
The failure is due to
the
``intrinsic exceptional point'' (IEP) features
of $H^{(IC)}$ and, in particular, to
the phenomenon of a
high-energy asymptotic parallelization
of its bound-state-mimicking eigenvectors.
In the paper it is argued that the operator $H^{(IC)}$
(and the like)
can only be interpreted
as a
manifestly unphysical, singular IEP
limit
of a hypothetical one-parametric family of certain
standard
quantum Hamiltonians.
For explanation,
an ample use is made of
perturbation theory and of multiple
analogies between IEPs and conventional Kato's
exceptional points.

\subsection*{Keywords}
.

quantum mechanics of unitary systems;

quasi-Hermitian Hamiltonians;

vicinity of intrinsic exceptional points;

amended perturbation theory;

imaginary cubic oscillator example;

 \newpage

\section{Introduction}

The concept of the so called ``intrinsic
exceptional point'' (IEP)
has been introduced by
Siegl and Krej\v{c}i\v{r}\'{\i}k who,
in their paper \cite{Siegl},
studied the ``prominent''
imaginary cubic (IC)
Schr\"{o}dinger equation
  \be
 H^{(IC)}\,|\psi_n^{(IC)}\kt = E_n^{(IC)}\,|\psi_n^{(IC)}\kt\,,
 \ \ \ n=0,1,\ldots\,
 ,
 \ \ \ H^{(IC)} = -\frac{d^2}{dx^2} + {\rm i}x^3\,.
 \label{SE}
 \ee
They felt motivated by
the
instability of the IC spectrum
under perturbations
\cite{Trefethen}.
They were able to complement such a
numerically supported
observation
by several rigorous mathematical
proofs (cf. also \cite{Viola}).
They found that
``the eigenvectors of the imaginary cubic oscillator do
not form a Riesz basis'' \cite{Siegl}.
In spite of having spectrum
which is real, discrete and bounded below \cite{DB,BB,DDT},
the manifestly non-Hermitian
IC Hamiltonian appeared not to be,
in the Riesz-basis sense,
diagonalizable.

Siegl with Krej\v{c}i\v{r}\'{\i}k
concluded that
``there is no quantum-mechanical
Hamiltonian associated with
it'' \cite{Siegl}.
The same authors
also
recalled the
standard mathematical terminology
and they reformulated their conclusion:
``In the language of exceptional points, the imaginary
cubic oscillator possesses an `intrinsic exceptional point' ''
which is,
as a singularity, ``much stronger
than any exceptional point associated
with finite Jordan blocks'' \cite{Siegl}.

These words are truly challenging,
having also motivated our study
of the role
of IEPs in
the deepest conceptual foundations of the contemporary
quantum physics.
It makes sense to
add that
Siegl with Krej\v{c}i\v{r}\'{\i}k
only
introduced the concept
via the above-cited remark, i.e.,
without giving a formal definition.
They 
specified IEP as an $N=\infty$ descendant of 
the conventional exceptional point of order $N$ (EPN, \cite{Kato}).
In this sense
the linear
{}{IEP}
differential operator $H^{(IC)}$ is really
``essentially different with respect to self-adjoint Hamiltonians''
\cite{Siegl}.

The problem of interpretation
of all of the non-standard, IEP-related
quantum bound-state problems remains,
at present, open.
In our contribution
to the currently running discussion
of this topic (cf. also \cite{UweS})
 {}{we will study and describe, more deeply,
the parallels as well as
differences between the two (viz., {the} IEP and EPN) concepts.}

We will start,
in section \ref{precedel}, from
a brief account of what is known about
the
linear-algebraic EPN analogues of
the ordinary differential IEP Eq.~(\ref{SE}).
{}{We will consider a
class of
Hamiltonians
({depending} on a real or complex parameter $g$)
which
admit a singular EPN
limit when $g \to g^{(EPN)}$.}
We will recall
{}{a few}
recent results
of the studies of
{}{this problem in which a suitable}
parameter-dependent $N$ by $N$ matrix
quantum Hamiltonian $H^{(N)}(g)$ {}{is considered}
at {}{a}
finite $N < \infty$.
We will emphasize the possibility and relevance of
{}{its}
canonical representation by
an $N$ by $N$ matrix
Jordan block {}{when $g \to g^{(EPN)}$}.

In the latter limit,
operator $H^{(N)}(g)$ 
ceases to be
diagonalizable
and, hence,
it ceases to be
acceptable as an eligible quantum Hamiltonian.
In section \ref{pocedel} we will emphasize that
many of its properties become really reminiscent
of the IEP features of the
differential-operator model (\ref{SE}) where
the corresponding Hilbert space of states is infinite-dimensional,
$N=\infty$.
We will
remind the readers
of the existing results concerning
physical meaning and impact of the
EPN-related finite-dimensional models.
We will
explain that
in many (often called ``quasi-Hermitian'' \cite{Dieudonne,Geyer})
quantum models of such a type
the limiting transition $g \to g^{(EPN)}$
can be interpreted
as one of the most natural realizations
of a genuine quantum phase transition
(cf., e.g., the description of a class of exactly solvable
models of such a process in \cite{passage}).

In section \ref{umdel} the emphasis will be shifted
to the $N \to \infty$ scenarios and
to the existence of several very useful
analogies between both of the IEP and EPN singular extremes.
We will point out that in such a comparison
the key role
of a methodical guide may be expected to be
played by (possibly, amended) perturbation theory.
Interested readers will be recommended to
find a phenomenologically oriented inspiration
as well as many related technical details
in 
older paper \cite{UweE}. 
The authors studied there
a fairly realistic non-Hermitian Hamiltonian
describing an N-particle Bose-Einstein condensate
generated by a sink and a source
in interaction.
Using
a combination of several complementary numerical as well as analytic
and perturbation methods
they managed to
detect the presence
of the EPN singularities
in their model.
They also revealed and explained that
under small perturbations
these singularities
did unfold in a very specific manner.

{}{These}
results
appeared encouraging because,
as the {}{authors} mentioned,
the 
``further investigations''
of the EPN-related problems
``remain tasks for future research.''
In our present paper, we just
decided to follow the recommendation.
In sections number \ref{jinan} and \ref{nejinan}
we will, in particular,
address {}{the} main technical {}{challenge
and we will propose} 
an IEP-related generalization of the
well known
perturbation-theory-based description of a generic unitary quantum
system near its {}{IEP} singularity.
We will succeed in showing that
many known technical tricks used and tested near EPN at $N<\infty$
can immediately be transferred
to the 
quantum-dynamical scenarios
in which the generic Hamiltonian
$H^{}(g)$
lies very close
to {}{its}
IEP limiting extreme.

In our last two sections \ref{ladel} and \ref{bladel}
and also in the series of six brief Appendices
we will {}{finally} complement our
considerations
by several
quantum-physics-oriented contextual
remarks.

\section{Conventional exceptional points associated
with finite Jordan blocks\label{precedel}}

In \cite{Siegl} we read that
the existence of the IEP singularity
``does
not restrict to the particular Hamiltonian''
of Eq.~(\ref{SE})
so that some
``new directions in
physical interpretation'' of all of the analogous
non-Hermitian quantum models
have to be sought
``since their properties
are essentially different with respect to self-adjoint Hamiltonians''
\cite{Siegl}.

This makes the IC model
important as a genuine methodical
as well as conceptual challenge.
Here, we intend to propose and advocate
the idea that
the resolution of the problem could be
guided by {}{another,} EPN-related
``good basis'' problem and by the existence of
parallels between quantum systems near their respective
IEP and EPN singularities.

The study of these parallels could proceed in several
independent
directions (cf. the samples of some of them in \cite{UweS}
or in \cite{Semoradova}).
In what follows, we will explain that and how
one of these directions could make use of
perturbation-expansion techniques.

\subsection{The phenomenon of EPN degeneracy}

In review paper \cite{ali}
the very first word of Abstract
emphasizes that
every operator $H$
eligible as an observable Hamiltonian
of a unitary quantum system
in Schr\"{o}dinger picture \cite{Messiah}
must be diagonalizable.
For any specific one-parametric family of Hamiltonians $H(g)$
such a requirement is not satisfied
in the EPN limit $g \to g^{(EPN)}$.
Then,
the operator can consistently be treated as Hamiltonian
only when $g \neq g^{(EPN)}$.

In an opposite direction of argumentation
one could recall the existence of exactly solvable
quasi-Hermitian
$N$ by $N$ matrix
models $H^{(N)}(g)$ of paper \cite{passage}
for which there exists a
vicinity of $g^{(EPN)}$
({}{i.e., say,
a suitable compact
and simply connected real or complex open domain ${\cal D}$
which does not contain  $g^{(EPN)}$ of course}) inside which
the respective quantum system is
found to
admit the standard physical probabilistic
interpretation.
{}{For  $g \in {\cal D}$, the}
diagonalizability of Hamiltonians $H^{(N)}(g)$
then {}{implies} that
we may
construct all of the bound-state
solutions of the
so called time-independent Schr\"{o}dinger equation
 \be
 H^{(N)}({g})\, |\psi_n(g)\kt
 = |\psi_n(g)\kt\,{E}^{(N)}_n(g)\,,\ \ \ \ \
 n=0,1,\ldots,N-1\,.
 \label{wfealt}
 \ee
Now, whenever the dimension of the Hilbert
space of states is finite, $N < \infty$,
we may immediately notice that even
in the generic non-degenerate case, all of the
eigenvalues ${E}^{(N)}_n(g)$
and eigenvectors $|\psi_n(g)\kt$
with $g \in {\cal D}$
degenerate in the ultimate (albeit
manifestly unphysical) EPN limit,
 \be
  \lim_{g \to g^{(EPN)}}\,{E}^{(N)}_n(g)=E^{(EPN)}\,,
  \ \ \ \
  \lim_{g \to g^{(EPN)}}\,|\psi_n(g)\kt=|\Psi^{(EPN)}\kt\,,
  \ \ \ \
  n=0,1,\ldots,N-1\,.
 \ee
This leads to the following observations:

\begin{itemize}

\item{[1]}
for all of the ``acceptable'' $g \neq g^{(EPN)}$ lying
in {}{the ``physical'', unitarity-compatible}
vicinity of the EPN value, $g \in {\cal D}$,
the normalized eigenvectors $|\psi_n(g)\kt$ of $H^{(N)}(g)$
are getting almost parallel to each other.

\item{[2]}
at the ``unacceptable'' value of $g = g^{(EPN)} \notin {\cal D}$
their
set
ceases to serve
as a basis suitable, say, for
the purposes of perturbation theory.

\item{[3]}
at $g = g^{(EPN)}$
one can still construct a ``good basis''
composed of the single remaining (degenerate)
eigenvector $|\psi_0(g^{(EPN)})\kt=|\Psi_0\kt$
and of an $(N-1)-$plet of linearly independent associated vectors
$|\Psi_j\kt$ with $j=1,2,\ldots,N-1$.

\end{itemize}

\subsection{EPN and modified Schr\"{o}dinger equation}

The
latter ``good basis''
can be perceived as an $N-$plet of column vectors.
They may be arranged
into the following formal $N$ by $N$ matrix,
 \be
 \{
 |\Psi_0\kt,|\Psi_1\kt,\ldots,|\Psi_{N-1}\kt
 \} := R^{(EPN)}
 \label{redef}
 \ee
called, usually, transition matrix.
Thus, we may introduce the two-diagonal Jordan block
 \be
 J^{(N)}
 \left (\eta\right )=
 \left (
 \begin{array}{ccccc}
 \eta&1&0&\ldots&0\\
 0&\eta&1&\ddots&\vdots\\
 0&0&\eta&\ddots&0\\
 \vdots&\ddots&\ddots&\ddots&1\\
 0&\ldots&0&0&\eta
 \ea
 \right )
 \,
  \label{JBK}
 \ee
and define the transition matrix as
solution of the
following Schr\"{o}dinger-like equation
 \be
 H^{(N)}({g}^{(EPN)})\, {R}^{(EPN)}
 = {R}^{(EPN)}\,{J}^{(N)}(E^{(EPN)})\,.
 \label{fealt}
 \ee
Interested readers are recommended to find a constructive illustration
of the reconstruction of transition matrix
${R}^{(EPN)}$ from the Hamiltonian in \cite{4a5}
where the illustrative solvable Hamiltonians
were real
matrices which were tridiagonal and multiparametric:
At $N=2J$ one had
 \be
 H^{(2{J})}(a,b,\ldots,z)=\left [\begin {array}{cccccccc}
  2{J}-1&z&0&\ldots&&&&
  \\
  -z&\ddots&\ddots&\ddots&\vdots&&&
 \\
  0&\ddots&3&b&0&\ldots&&
 \\
 \vdots&\ddots&-b&1&a&0&\ldots&
 \\
 &\ldots&0&-a&-1&b&0&\ldots
  \\
 &&\ldots&0&-b &-3&\ddots&
 \\
 &&&\vdots&\ddots&\ddots &\ddots&z
 \\
 &&&&\ldots&0&-z&1-2{J}\end
 {array} \right ]
 \label{sudam}
 \ee
etc (for a few further
related comments see also Appendices A.1 and A.2 below).

\section{The mechanism of unfolding of the EPN degeneracy\label{pocedel}}

\subsection{{}{The} hypothesis of admissibility of
{}{at least some $g \approx {g}^{(EPN)}$}}

The {}{purpose} of the
above-outlined {}{choice} of the basis
is twofold. Firstly,
it enables us to re-read
our Schr\"{o}dinger-like Eq.~(\ref{fealt})
as an equivalent linear-algebraic relation
 \be
 \left [{R}^{(EPN)}\right ]^{-1}\,
 H({g}^{(EPN)})\, {R}^{(EPN)}=
 {J}^{(N)}({E}^{(EPN)})
 \,
  \label{refealt}
 \ee
i.e.,
as a definition of
a canonical
Jordan-block
representation
of any non-Hermitian  Hamiltonian of interest
at its EPN singularity.
Secondly,  the
amended basis
{}{will} find application
in a reformulation of
standard
perturbation theory.
In such a reformulation,
the
role of the unperturbed Hamiltonian
{}{will be}
played by
its unphysical, singular EPN limit.
The trick is that we
use the columns of ${R}^{(EPN)}$ as unperturbed basis.
In the overall perturbation-theory spirit,
the perturbed system acquires {}{then} a standard
phenomenological interpretation for the parameters $g$ lying inside
 a suitable ``physical''
vicinity  ${\cal D}$
of the EPN singularity.

The latter philosophy is to be advocated and used in what follows.
We will only assume the knowledge of the transition
matrix $ {R}^{(EPN)}$ at
an exceptional point of finite order
and we will {}{then} extend the use of this basis
to {}{a}
vicinity of the singularity.
This will enable us to
invert the limiting process $g \to {g}^{(EPN)}$
and to consider the original Hamiltonians at some
$g \neq {g}^{(EPN)}$.
Our knowledge of transition matrix
will yield the model
described as a perturbation of {}{the} Jordan
block matrix,
 \be
 \left [{R}^{(EPN)}\right ]^{-1}\,
 H^{(N)}({g})\, {R}^{(EPN)}=
 {J}^{(N)}({E}^{(EPN)}) +  \lambda\,V^{(N)}({g})\,.
  \label{berefealt}
 \ee
{\it A priori}, we will only have to demand that the auxiliary variable
$\lambda = {\cal O}({g}-{g}^{(EPN)})$ remains small.

\subsection{The possibility of keeping the perturbed spectrum real}

In papers \cite{admissible,corridors} we
considered the above-mentioned quantum-dynamics scenarios and
we
studied there
the criteria of smallness of the perturbations
$V^{(N)}$.
We showed that the conditions of the stability
and unitarity
of the system
can be given a
mathematically as well as phenomenologically consistent form.

For illustration 
let us set
$E^{(EPN)}=0$
and let us consider
the
bound-state problem
as a perturbation of its EPN limit,
    \be
 \left [
 {J}^{(N)}(0) + \lambda\,V^{(N)}
  \right ]
 \,|{\Psi}(\lambda)\kt=\epsilon(\lambda)\,|{\Psi}(\lambda)\kt\,.
  \label{perL1}
 \ee
{}{With} the energy levels counted, whenever needed, by a
subscript or superscript,
we will never use this index,
keeping it just dummy.
We will rather introduce another subscript which will
run, say, from $1$ to $N$ and which will
number the components $\Psi_j$ of the ket vector $|{\Psi}\kt$
(here we are also dropping the argument $\lambda$ as redundant).
This convention enables us to
fix
the norm of $|{\Psi}\kt$ by the choice of ${\Psi}_1=1$  and
to define another, ``shifted'' column vector
 \be
  \left (
 \ba
 {\Psi}_2\\
 {\Psi}_3\\
 \vdots\\
 {\Psi}_{N}\\
 \Omega_{N}
 \ea
 \right ):=\left (
 \ba
 y_1\\
 y_2\\
 \vdots\\
 y_{N-1}\\
 y_{N}
 \ea
 \right )=\vec{y} \,
 \label{jtarov}
 \ee
where $\Omega_N$ is a new auxiliary variable.
Next we notice that the $N$ by $N$
matrix
  \be
 A=A(N,\epsilon)=\left (
 \begin{array}{ccccc}
 1&0&0&\ldots&0\\
 \epsilon&1&0&\ddots&\vdots\\
 \epsilon^2&\epsilon&\ddots&\ddots&0\\
 \vdots&\ddots&\ddots&1&0\\
 \epsilon^{N-1}&\ldots&\epsilon^2&\epsilon&1
 \ea
 \right )\,
 \label{9}
 \ee
is just an inverse of
two-diagonal matrix
 \be
 A^{-1}
 =\left (
 \begin{array}{ccccc}
 1&0&0&\ldots&0\\
 -\epsilon&1&0&\ddots&\vdots\\
 0&-\epsilon&\ddots&\ddots&0\\
 \vdots&\ddots&\ddots&1&0\\
 0&\ldots&0&-\epsilon&1
 \ea
 \right )\,.
 \label{9inv}
  \ee
Finally we select
the first
column of the matrix in Eq.~(\ref{perL1})
and {}{we} denote it by another dedicated symbol,
 \be
  \left (
 \ba
 \epsilon - \lambda\,{V}_{1,1}\\
 - \lambda\,{V}_{2,1}\\
 \vdots\\
 - \lambda\,{V}_{N,1}
 \ea
 \right ):=
 \vec{r} = \vec{r}(\lambda) \,.
 \label{postaru}
 \ee
All of these abbreviations
convert our initial homogeneous Schr\"{o}dinger
Eq.~(\ref{perL1}) into its equivalent matrix form
 \be
 (A^{-1} + \lambda\,Z)\, \vec{y}= \vec{r}\,
 \label{tarov}
 \ee
or, better,
 \be
 (I + \lambda\,A\,Z)\, \vec{y}= A\,\vec{r}\,
 \label{satarov}
 \ee
where the symbol $Z$ stands
for a modified form of the
matrix of perturbation,
 \be
 V^{(N)} \ \to \ Z
 =\left (
 \begin{array}{ccccc}
 {V}_{1,2}&{V}_{1,3}&\ldots&{V}_{1,N}&0\\
 {V}_{2,2}&{V}_{2,3}&\ldots&{V}_{2,N}&0\\
 \ldots&\ldots&\ldots&\vdots&\vdots\\
 {V}_{N,2}&{V}_{N,3}&\ldots&{V}_{N,N}&0
 \ea
 \right )\,.
 \label{ustahromu}
 \ee
In paper \cite{admissible}
we proved that the construction of the
perturbation corrections now becomes reduced to 
self-consistency
condition
 \be
 \Omega_N=0\,.
 \label{compat}
 \ee
In the
same reference, interested readers may also find
an explicit form of the construction in the
leading-order
approximation.

{}{Its} basic aspects are worth recalling
because they immediately
help {}{us} to
clarify
the meaning of
the rather vague
assumption of the smallness of 
perturbation. {}{It} is sufficient to
employ the Taylor-series
expansion of the resolvent
which yields formula
 \be
 \vec{y}^{(solution)}(\epsilon)= A\,\vec{r}-
 \lambda \,A\,\,Z\,
 A\,\vec{r}+\lambda^2 \,A\,\,Z\,A\,Z\,
 A\,\vec{r}
 -\ldots\,.
 \label{tadef}
 \ee
Such a
wave-function-representing ket-vector
depends on the variable
parameter
$\epsilon$  but, ultimately, all of the eligible values of
$\epsilon$  become
fixed by constraint (\ref{compat}).

The latter constraint plays the role of
secular equation which has
the
{}{single} vector-component form
 \be
 {y_N}^{(solution)}(\epsilon)=0\,.
 \label{krutadef}
 \ee
In the last step of the construction
we 
have to solve such
an explicit transcendental equation
in order to get
all of the alternative perturbation-generated energy
corrections $\epsilon$.
In a direct dependence on the model in question,
precisely the study of the roots of this equation also
offers the criterion
of the reality of the whole spectrum
in the leading-order approximation.

\section{{}{Large $N$} and anomalous Hamiltonians\label{umdel}}

The message to be
extracted from the
above-outlined EPN-based construction
is that
for the purposes of transition to its IEP analogue
we may try to make use of the
IEP - EPN similarities.
The main one will consist in the
unperturbed-Hamiltonian
interpretation
of the singular IEP operator
tractable, in some sense,
as a large-$N$ descendant of its
finite-$N$
EPN analogues.

In the analysis of {}{both of the EPN and} IEP singularities
a central role is certainly played
by the phenomenon of the asymptotic confluence
of {}{finitely or} infinitely many eigenvectors,
not accompanied
by the confluence
of the eigenvalues
{}{in the IEP case. This can be found confirmed}
in
\cite{UweS} where
we read that 
``for matrices approaching an
exceptional point, it is
known \cite{[28]}
that the corresponding eigenvectors are tending
to coalesce. For the infinite-dimensional Hilbert space
(and Krein space) setup of the IC model, the
eigenfunctions of the Hamiltonian having diverging projector
norms and asymptotically approaching a {\cal PT} phase
transition region at spectral infinity signal a possible tendency
toward collinearity and isotropy of an infinite number
of these eigenfunctions''
\cite{UweS}.

\subsection{The {}{phenomenon of} asymptotic degeneracy of eigenvectors}

The
EPN - IEP parallels are certainly incomplete.
{}{Still, in both cases}
an amendment of the notation
might prove useful.
{}{Here, we will follow the
notation {}{convention which was} proposed in
our comprehensive review paper \cite{SIGMA}.
In this spirit, the}
first mathematical subtlety which we will
have to keep in mind is that for a generic IEP model
the spectrum itself remains non-degenerate. {}{Still,}
in a way sampled by the IC example, the generic
IEP  Schr\"{o}dinger equation
  \be
 H^{(IEP)}\,|\psi_n^{(IEP)}\kt = E_n^{(IEP)}\,|\psi_n^{(IEP)}\kt\,,
 \ \ \ n=0,1,\ldots\,
 \label{aSE}
 \ee
can  be considered analogous to its
{}{finite-dimensional EPN-supporting} partners.

Once the spectrum is found real and discrete
(which is precisely the case
of {}{our} illustrative IC Schr\"{o}dinger
Eq.~(\ref{SE})),
the same property also
characterizes the
formally independent
Hermitian conjugate
Schr\"{o}dinger equation problem
  \be
 \left [H^{(IEP)}
 \right ]^\dagger\,|\psi_n^{(IEP)}\kkt = E_n^{(IEP)}\,|\psi_n^{(IEP)}\kkt\,,
 \ \ \ n=0,1,\ldots\,.
 \label{aSEb}
 \ee
{}{Here, our use of the ``ketket'' symbol $ \kkt $
deserves an immediate comment and explanation. Mainly
because it is closely connected
with its role played in
the
three-Hilbert-space reformulation of the conventional quantum mechanics
of unitary systems
as described, e.g., in review paper \cite{SIGMA}.
For the reasons explained in
the three dedicated Appendices A. 4 -- A. 6 below,
the latter formalism is also -- implicitly --
recalled and used in
our present paper.
In these Appendices,
interested readers may find a
more extensive commentary on the entirely
equivalent three-Hilbert-space version
of the standard textbook quantum theory, with more emphasis put upon
some questions of the physical probabilistic interpretation
of the illustrative physical models of our present methodical interest.}

The IEP-characterizing phenomenon
of
asymptotic degeneracy
enables us to re-establish the above-mentioned analogy
with the EPN form of confluence of the  eigenfunctions.
This phenomenon involves, first of all, the right
eigenvectors $|\psi_{n}^{(IEP)}\kt$ of $H^{(IEP)}$.
For them we have
 \be
 |\psi_{M+k}^{(IEP)}\kt \approx |\psi_{M+k+1}^{(IEP)}\kt\,,
 \ \ \ \ k=1,2,\ldots\,
 \label{iota}
 \ee
at  $M\gg 1$.
Similarly,
the degeneracy concerns also
the left {}{eigenvectors}
{\it alias\,} ``brabra''
eigenvectors $\bbr \psi_{n}^{(IEP)}|$ of
the same {}{non-Hermitian operator} $H^{(IEP)}$.
Often we rather refer to their
conjugate form $| \psi_{n}^{(IEP)}\kkt$ of ``ketket''
eigenvectors of conjugate $\left [H^{(IEP)}\right ]^\dagger$.
In this representation we {}{encounter an entirely analogous
IEP-related confluence of the eigenvectors,}
 \be
 |\psi_{M+k}^{(IEP)}\kkt \approx |\psi_{M+k+1}^{(IEP)}\kkt\,,
 \ \ \ \ k=1,2,\ldots\,.
 \label{bota}
 \ee
In both Eqs.~(\ref{iota}) and (\ref{bota})
the degree of confluence
depends on the Hamiltonian and
it may be expected to grow with the growth of $M$.

The phenomenon
of the confluences (\ref{iota}) and (\ref{bota})
finds its formal predecessor
in the finite-dimensional case in which,
during the transition to
singularity $g \to g^{(EPN)}$, the elements of
the $N-$plet of eigenvectors
of
any preselected $N$ by $N$ Hamiltonian
matrix $H=H^{(N)}(g)$ really
lose their mutual linear independence.
Still, the analogy of a genuine IEP system
with the IEP-mimicking $N=\infty$ EPN extreme
is incomplete since in the former case
the spectrum
remains non-degenerate.
A more explicit analysis is necessary.

\subsection{Canonical representation  of $H^{(IEP)}$\label{ggaone}}

The IEP-characterizing non-degeneracy of eigenvalues
can be perceived as a serendipitious simplification
of their study. Still,
a decisive
IEP-related difficulty results from the
effective
asymptotic parallelization of the unlimited
number of eigenvectors.

This forces us to recall,
as our main source of inspiration,
relations (\ref{JBK}) and (\ref{fealt})
of section \ref{precedel} above.
In the IEP case
our key task can be now identified as
an appropriate generalization of the
transition matrices ${R}^{(EPN)}$ which played key role
in the perturbation-theory considerations of section \ref{pocedel}.
In other words, we have to replace Eq.~(\ref{fealt})
by a  modified
eigenvalue-like problem
 \be
 H^{(IEP)}\, {\cal R}^{(IEP)}
 = {\cal R}^{(IEP)}\,{\cal J}^{(IEP)}\,
 \label{mfealt}
 \ee
in which the low-lying eigenstates
do not require any specific attention.
Thus, the conventional Jordan-block-like
bidiagonal (i.e., minimally non-diagonal) canonical-matrix
structure of Eq.~(\ref{JBK})
will only reemerge here in a
infinite-dimensional submatrix of
upgraded
 \be
 {\cal J}^{(IEP)}=
  \left (
 \begin{array}{cccc|cccc}
 E_0&0&\ldots&0&0&\ldots&&\\
 0&E_1&\ddots&\vdots&\vdots&&&\\
 \vdots&\ddots&\ddots&0&0&\ldots&&\\
 0&\ldots&0&E_{K-1}&0&0&\ldots&\\
 \hline
 0&\ldots&0&0&E_{K}&1&0&\ldots\\
 \vdots&&\vdots&0&0&E_{K+1}&1&\ddots\\
 &&&\vdots&0&0&E_{K+2}&\ddots\\
 &&&&\vdots&\vdots&\ddots&\ddots\\
 \ea
 \right )
 \,.
  \label{mJBK}
 \ee
In this arrangement
the partitioning of the basis
may be characterized by
the projectors $P$ (on the first $K$ lowest eigenstates {}{of $H^{(IEP)}$})
and {}{$Q$
(such that the unit operator $I$ in Hilbert space can be decomposed as follows,
$I = Q+P$).}

In a certain parallel with EPN,
a key technical {}{step will
now be a suitable perturbation-mediated}
weakening or removal of the asymptotic degeneracies
(\ref{iota})
and (\ref{bota})
{}{of the asymptotic
eigenstates of $H^{(IEP)}$}.

\section{Towards a regularization of $H^{(IEP)}$s by perturbation\label{jinan}}

From a purely historical point of view
the idea of ``prominence'' of the IEP-sampling
Schr\"{o}dinger Eq.~(\ref{SE})
dates back to
its methodical role in field theory \cite{Fisher}
and to the Bessis' and Zinn-Justin's
empirically revealed
conjecture
(cf. \cite{DB}, cited also in \cite{BB})
that the spectrum $\{E_n^{(IC)}\}$ of $H^{(IC)}$
is real, discrete and bounded from below,
i.e., tractable, in principle at least,
as {}{a} set of
observable energy levels.
In spite of the manifest non-Hermiticity
of Hamiltonian $H^{(IC)}$,
the model was temporarily
accepted
as potentially compatible with all of the principles and
postulates of quantum mechanics. The
corresponding technical details may be found
in
review paper \cite{Carl}.

Unfortunately,
the end of the excitement came after the
Siegl's and Krej\v{c}i\v{r}\'{\i}k's
rigorous proof
that the IC model cannot in fact be assigned
any form of conventional probabilistic interpretation
in
a mathematically
consistent manner~\cite{Siegl}.
More or less the same conclusion has been also made, by
G\"{u}nther and Stefani,
in a not yet published preprint \cite{UweS}.
At present,
in the context of the unitary-evolution part of
non-Hermitian
quantum mechanics
the problem
of
a correct physical interpretation
of the IC model
{}{itself}
remains
unresolved.

Concerning the future developments, we remain a bit skeptical because
the IEP nature of the IC model looks, in many a respect,
only too similar to its
much better understood (and manifestly singular and
unphysical) EPN-related finite-dimensional analogues.

\subsection{The IEP - EPN differences and parallels}

The essence of the anomalous nature of any IEP-related
Hamiltonian $H^{(IEP)}$ lies in the
asymptotic degeneracies (\ref{iota}) and (\ref{bota})
of its respective right and left eigenvectors.
At the same time,
a weak point of the
amendment of the basis as mediated by the choice of non-diagonal matrix (\ref{mJBK})
may be seen in the necessity of specification of a
``sufficiently large'' onset $K\gg 1$ of the
de-parallelization.
Such a specification
is just numerically, computer-precision
motivated. In contrast to the above-outlined treatment of the
EPN scenarios where the dimension $N$ was fixed,
the IEP-implied choice of any finite $K$ is
purely pragmatic, {}{immanently approximative} and
virtually arbitrary.

We now intend to show that, surprisingly enough,
the apparently more or less accidental
flexibility of our choice of $K$
can in fact become an important {}{mathematical} tool
facilitating an EPN-resembling regularization and
consistent interpretation
of quantum systems near their
IEP singular extreme.

First of all, there is no doubt about
the necessity of transition from the less suitable
unperturbed
basis (composed of eigenvectors)
to an ``anomalous''
basis resembling Eq.~(\ref{redef}).
The reason is {}{provided}
by Eqs.~(\ref{mfealt}) and (\ref{mJBK}):
only
a rectification of the
underlying biorthogonal or biorthonormal basis \cite{Brody}
can re-establish the  EPN - IEP parallels
even when
achieved, also in the latter case,
at {}{the} expense of
non-diagonality
and non-Hermiticity of matrix (\ref{mJBK}).

Although the EPN
singularity
encountered at finite matrix dimensions $N<\infty$
is,
according to paper \cite{Siegl},
perceivably weaker
than its IEP
analogue,
the essence of our present message
will be complementary. Basically, we will
emphasize that
one can also reveal and make a productive use of
certain partial similarities
between the two concepts.
In particular, we propose that the above-outlined
possibility and feasibility of treating
the manifestly unphysical finite-dimensional singular matrices
$H^{(N)}(g^{(EPN)})$
as formally acceptable
unperturbed Hamiltonians
is to be transferred also to the IEP context.
An {}{anomalous} ``good''
basis composed of the
columns of transition matrix
should be, {\it mutatis mutandis},
reconstructed also from any given
Hamiltonian $H^{(IEP)}$.

\subsection{IEP-unfolding bases}

The
IEP (i.e., $N=\infty$) and
EPN (i.e., $N<\infty$) singularities share
the phenomenon of
the parallelization of eigenvectors.
In a small vicinity of the singularity
the analysis has to rely
upon a properly adapted form of
perturbation-theory.
Our present proposal of transfer of this idea from EPN to IEP
will be inspired, therefore,
by section \ref{pocedel}.

The parallels are, naturally, incomplete so that in the IEP setting
certain truly specific features have to be expected to
emerge.
For the purposes of clarification
let us mention that even if we fix a
finite $K\gg 1$
it remains far from obvious how to
follow the analogy with Eqs. (\ref{wfealt}) and (\ref{redef})
and how to treat also $H^{(IEP)}$
as an unperturbed Hamiltonian.
The reason is that
we do not have any immediate
analogue of Eq.~(\ref{berefealt}).
In the models as sampled by the IC oscillator
we also miss a
parameter $g$ or $\lambda$ which would
control the form and size of perturbations
needed for a phenomenologically
motivated  unfolding of the
manifestly unphysical IEP singularity.

This being admitted,
we may still be guided by the EPN dynamical scenario
as outlined in the preceding sections \ref{precedel} and  \ref{pocedel}.
In the study of the IEP systems, first of all,
we should construct a good
unperturbed basis in Hilbert space, therefore.
The most natural IEP analogue
of the EPN-related Jordan-block-matrix (\ref{JBK})
is to be seen in its IEP-related amendment (\ref{mJBK}),
rendering
the EPN-related
unperturbed Schr\"{o}dinger-like Eq.~(\ref{fealt})
replaced by its IEP-related alternative (\ref{mfealt}).

In connection with the standard
and unmodified conjugate eigenvalue problems
(\ref{aSE}) and (\ref{aSEb})
the difficulty is that
in the Siegl's and Krej\v{c}i\v{r}\'{\i}k's words
``the eigenvectors, despite possibly being complete,
do not form a `good' basis,
i.e., an unconditional (Riesz) basis''
\cite{Siegl}.
Thus, the left and right eigenvectors of $H^{(IEP)}$
can only be used as a basis in the $P-$projected subspace of the
Hilbert space.
Otherwise, the
EPN - IEP parallelism has to be fully taken into account, i.e.,
in Eq.~(\ref{mfealt}),
one has to recall the EPN-related definition
(\ref{redef})
and
define, in Eq.~(\ref{mfealt}), its present IEP-related
calligraphic-symbol partner
${\cal R}^{(IEP)}$
as the following concatenated infinite
set of column vectors
 \be
 {\cal R}^{(IEP)}=
 \{
 |\psi_0\kt,|\psi_1\kt,\ldots,|\psi_{K-1}\kt,
 |f_{K}\kt,
 |f_{K+1}\kt,
 |f_{K+2}\kt,\,
 \ldots
 \}\,.
 \label{uredef}
 \ee
{}{This array is} composed of the mere first $K$
eigenkets $|\psi_j\kt$
complemented by the modified, associated-like ket vectors
$|f_{K+k}\kt$ with $k=0,1,2,\ldots$.

\subsection{Recurrences}

The possibility (and also, in some sense, the necessity)
of the explicit construction of the latter subfamily of
the new ket vectors
is in fact the very core of
our present innovation
of the foundations of
the formalism of quantum mechanics.
Briefly, our basic
message is that
in a way which parallels the EPN-related
considerations of section \ref{pocedel}
above,
our present introduction
of the nontrivial
IEP-motivated transition matrix (\ref{uredef})
may be expected to play a key role in
the regularization of any singular $H^{IEP)}$
via its suitable small perturbations.

The
replacement of eigenvectors
$|\psi_{K+k}\kt$ by non-eigenvectors
$|f_{K+k}\kt$ in (\ref{uredef})
has to
weaken the asymptotically increasing
parallelism between the subsequent columns of
the transition matrix ${\cal R}^{(IEP)}$.
The
infinite-dimensional matrix form of
transition matrix (\ref{uredef})
makes this task different from its EPN predecessor.
In technical terms, the insertion of
(\ref{uredef}) may be used to reduce
the nontrivial part of Eq.~(\ref{mfealt})
to the sequence of recurrences
 \be
 \left (H^{(IEP)}-E^{(IEP)}_{K+m} \right)\,|f_{K+m}\kt=|f_{K+m-1}\kt
 \,,\ \ \ \
  m=1,2,\ldots\,
 \label{serel}
 \ee
with the initial choice of $|f_{K}\kt=c_{0,0}|\psi_{K}\kt$
using any $c_{0,0}\neq 0$.

The solution of these relations
can be then given the form of finite sum
 \be
 |f_{K+p}\kt= \sum_{n=0}^p \,c_{p,n}\,|\psi_{K+n}\kt \,,
 \ \ \ \ p=0,1,\ldots
 \label{celer}
  \ee
where the leading coefficient $c_{k,k}$
is arbitrary.
Now we assume and recall the biorthonormality of the eigenbasis yielding
$\bbr \psi_m|\psi_n\kt = \delta_{m,n}$ and
enabling us to convert Eq.~(\ref{serel}),
i.e., the
recurrences for kets
into the recurrences for coefficients,
 \be
 c_{k,m}= (E_{K+m}-E_{K+k})^{-1} c_{k-1,m}\,,
 \ \ \ \ \ m = 0,1,\ldots,k-1\,,
 \ \ \ \ \ k=1,2,\ldots\,.
 \label{finrek}
 \ee
Our freedom of choice of
the highest-component
coefficients $c_{k,k}$
enables us to suppress
the IEP-accompanying asymptotic
parallelization of the vectors of basis in Hilbert space.
The goal is achieved.
For every
particular IEP model we {}{may} recall recurrences
(\ref{finrek}) and replace the
$Q-$projected part of the
basis composed of
eigenvectors by the
$Q-$projected part of the basis composed, up to the first item
$|f_{K}\kt=c_{0,0}|\psi_{K}\kt$, of
non-eigenvectors. And
this is precisely what has been done in Eq.~(\ref{uredef}).

\section{Constructive IEP-perturbation considerations\label{nejinan}}

\subsection{Formulation of the problem}

G\"{u}nther with Stefani \cite{UweS}
stated that
``what is still lacking''
in
the IEP setup
``is a simple physical explanation
scheme for the non-Rieszian behavior of the
eigenfunction sets''.
We agree. We are even more skeptical
because we would rather say that the expected
`simple physical explanation'
making, in particular, the
popular IC oscillators (\ref{SE}) physical
need not exist at all. Indeed,
we believe that
a consistent physical
closed quantum system
interpretation
could much more easily
be assigned to
suitable
perturbations of the
``seed'' IEP oscillators
with uncertain interpretation
(cf. {}{also} \cite{Mityagin} in this respect).

Our belief is supported by the
existence of parallels between the IEP and EPN scenarios.
On this ground one could really become able to assign a
sound phenomenological meaning
to many hypothetical
parameter-dependent Hamiltonians
${\cal H}^{(new)}({\lambda})$
defined as
certain ``admissible'' {perturbations}
of the extreme IEP
reference operators
 $$
 H^{(IEP)}
 \, \equiv \,{\cal H}^{(new)}({0})
 $$
(see, in this respect, also the methodical
guidance as provided by the
illustrative EPN-related
Eq.~(\ref{uho})
in section \ref{ladel} below).

Open
questions emerge when we
fix a sufficiently large $K$,
separate the Hilbert space of states into its two
more or less decoupled subspaces and when we finally
introduce a hypothetical perturbed
Hamiltonian ${\cal H}^{(new)}(\lambda)$ and the following IEP analogue
of
Eq.~(\ref{berefealt}),
 \be
 \left [{\cal R}^{(IEP)}\right ]^{-1}\,
 {\cal H}^{(new)}({\lambda})\, {\cal R}^{(IEP)}=
 {\cal J}^{(IEP)} +  \lambda\,{\cal V}
 \,.
  \label{umberefe}
 \ee
The analogy with EPNs
is incomplete because here,
the spectrum of
the unperturbed zero-order Hamiltonian
remains {\em non-degenerate}.
This is a simplification which can be perceived
as partially compensating
the increase of the overall complexity of the IEP problem.

Incidentally, a similar simplification has also been detected
in the
realistic EPN-supporting Bose-Hubbard
model of paper  \cite{UweE} where,
in dependence on parameters,
the authors had to use
{\em both\,} the degenerate and non-degenerate
versions of perturbation theory.
Thus,
no abstract conceptual problems
have to be expected to emerge after one returns to the generic
IEP-related dynamics.
Still, as long as the IEP-related problems are infinite-dimensional,
the perturbed IEP spectrum
cannot be deduced from any analogue
of the EPN-based implicit-definition
constraint (\ref{compat}).
The study of properties of the
vicinity of the IEP singularity
cannot be based on a direct
reference to its EPN
analogue.
{}{The methods of construction
have to be amended.}

\subsection{Structure of solutions}

In the light of the $P + Q$ partitioning
of matrix ${\cal J}^{(IEP)}$ {}{in} Eq.~(\ref{mJBK})
the constructiton of the perturbed forms of
the
low-lying bound states
remains standard.
The $P-$projected states may
be ignored just
as certain decoupled observers.
Only the treatment of the ``asymptotic'',
$Q-$projected
components of the quantum system in question
becomes difficult and singular,
``for instance, due to spectral instabilities'' \cite{Siegl}.

This leads to
the necessity of solving the perturbed
Schr\"{o}dinger equation
 \be
 \left[{\cal J}^{(IEP)} +  \lambda\,{\cal V}
 \right]\,|\psi(\lambda)\kt
  =E(\lambda)\,|\psi(\lambda)\kt
 \ee
where $\lambda \neq 0$
(so that we avoid the IEP singularity)
and
where we have to insert
 \be
 |\psi(\lambda)\kt
 =|\psi(0)\kt +\lambda\,
 |\psi^{[1]}\kt+\lambda^2\,
 |\psi^{[2]}\kt+\ldots
 \label{[33b]}
 \ee
and
 \be
 E(\lambda)=
 E(0) +\lambda\,
 E^{[1]}+\lambda^2\,
 E^{[2]}\kt+\ldots\,.
 \label{[33]}
 \ee
In the light of definition (\ref{mJBK})
the $P-$projected
part
of our unperturbed
Hamiltonian
${\cal J}^{(IEP)}$
is a diagonal matrix
containing the unperturbed bound state energy eigenvalues
$E_0$,
$E_1$,
\ldots
$E_{K}$.
All of the related perturbed low-lying bound states
can be then constructed using the conventional
Rayleigh-Schr\"{o}dinger
perturbation theory of
textbooks \cite{Messiah}.
For any practical purposes
it is fully acceptable just to
make the choice of a sufficiently large
dimension $K$  of the $P-$projected subspace, therefore.

In our present, conceptually more ambitious analysis of the
problem it makes sense to turn attention to
the states with the high-lying zero-order
unperturbed energies.
The related unperturbed ket vectors
$|\psi(0)\kt$
will
lie in the
complementary (and infinite-dimensional)
$Q-$projected subspace of Hilbert space.
The more or less conventional construction
of its perturbed descendant {}{given by} Eq.~(\ref{[33b]})
will then possess several anomalous features
of course.

The first anomaly is that the $Q-$projected
part
$Q{\cal J}^{(IEP)}Q$
of our unperturbed
Hamiltonian
is manifestly non-Hermitian.
Even after a tentative
finite-matrix
truncation of the perturbed eigenvalue problem
using a sufficiently large
cut-off $N \gg K \gg 1$ of the Hilbert space bases,
the implementation of the conventional
Rayleigh-Schr\"{o}dinger recipe
would require a spectral representation
of the unperturbed Hamiltonian operator.

Due to the specific upper-triangular two-diagonal
structure of matrix
$Q{\cal J}^{(IEP)}Q$,
the construction of a biorthonormalized basis would be needed.
Thus, the left eigenvectors of $Q{\cal J}^{(IEP)}Q$ (i.e.,
in the notation of paper \cite{SIGMA}, ketkets, $|\chi_j\kkt$)
will be
complicated and different from their
right-eigenvector biorthogonal partners $|\chi_j\kt$.
This means that also the conventional
Rayleigh-Schr\"{o}dinger
elementary unperturbed projectors {}{$|\chi_j\kt \bbr\chi_j| $}
(needed during the construction)
will have a practically prohibitively complicated
explicit matrix structure.

One could also find another, more immediate
indication of the possible emergence of
irregularities in section \ref{pocedel}
where Eq.~(\ref{krutadef}) playing the role of
an ultimate
transcendental equation
determining all of the perturbed EPN eigenvalues
was just a constraint imposed upon the
very last,
$N-$th component of a relevant ket vector.
Needless to add that in the IEP
setting one should consider $ N \to \infty$
so that the direct analogy with EPNs gets broken.

Another, independent word of warning might originate from
the fact that
for all of the truly high
energy levels $E_{K+m}(\lambda)$ with $m\gg 1$
the use of the
explicit Rayleigh-Schr\"{o}dinger recipe
would require the derivation of formulae which
would be $m-$dependent and
different for the different, {}{i.e., for the $(K+m)-$th,}
$m-$numbered
excitations.
Fortunately, the latter technical obstacle and
difficulty has
a comparatively elementary resolution
because
what is fully at our disposal is
our choice of the value of $K$.
We may feel free to work, exclusively, with
the properly innovated Rayleigh-Schr\"{o}dinger formulae
deduced just
at a single value of $m$, i.e., say, at $m=0$.

This certainly simplifies our task. In
methodical setting,
it will be sufficient
to
work with the Hamiltonian of Eq.~(\ref{mJBK}) at $K=0$.
Thus,
one just has to solve
Schr\"{o}dinger equation
\be
 \hspace{-1.6cm}\left [ \left (
 \begin{array}{cccc}
 E_{0}-E(\lambda)&1&0&\ldots\\
 0&E_{1}-E(\lambda)&1&\ddots\\
 0&0&E_{2}-E(\lambda)&\ddots\\
 \vdots&\ddots&\ddots&\ddots\\
 \ea
 \right )
 +\lambda\,
 \left (
 \begin{array}{cccc}
 {\cal V}_{00}&{\cal V}_{01}&{\cal V}_{02}&\ldots\\
 {\cal V}_{10}&{\cal V}_{11}&{\cal V}_{12}&\ddots\\
 {\cal V}_{201}&{\cal V}_{21}&{\cal V}_{22}&\ddots\\
 \vdots&\ddots&\ddots&\ddots\\
 \ea
 \right )\right ]\,|\psi(\lambda)\kt
  =0
 \,
  \label{permJBK}
 \ee
where
\be
 |\psi(\lambda)\kt
 =
 \left (
 \begin{array}{c}
 \psi_0(0)\\
 \psi_1(0)\\
 \vdots
 \ea
 \right )
 +
 \sum_{k=1}^{\infty}\lambda^k\,
 \left (
 \begin{array}{c}
 \psi_0^{[k]}\\
 \psi_1^{[k]}\\
 \vdots
 \ea
 \right )\,.
 \ee
As long as
$E_{0}=E(0)$ and $|\psi_j(0)\kt = 0$ at all $j\neq 0$,
it makes sense to
abbreviate $E_k-E(\lambda):=F_k(\lambda)$
and remember that $F_0(\lambda)=\lambda E^{[1]}+
{\cal O}(\lambda^2)$.

In the first-order
approximation we have, therefore, equation
 \be
  \left (
 \begin{array}{cccc}
 -\lambda E^{[1]}&1&0&\ldots\\
 0&F_1(0)&1&\ddots\\
 0&0&F_2(0)&\ddots\\
 \vdots&\ddots&\ddots&\ddots\\
 \ea
 \right )
 \left [
 \left (
 \begin{array}{c}
 \psi_0(0)\\
 0\\
 0\\
 \vdots
 \ea
 \right )
 +
 \lambda\,
 \left (
 \begin{array}{c}
 \psi_0^{[1]}\\
 \psi_1^{[1]}\\
 \psi_2^{[1]}\\
 \vdots
 \ea
 \right )
 \right ] =
  \ee
 \be
 =\lambda\,
 \left (
 \begin{array}{c}
 -E^{[1]}\,\psi_0(0)\\
 0\\
 0\\
 \vdots
 \ea
 \right )
 +
 \lambda\,
  \left (
 \begin{array}{cccc}
 1&0&o&\ldots\\
 F_1(0)&1&0&\ddots\\
 0&F_2(0)&1&\ddots\\
 \vdots&\ddots&\ddots&\ddots\\
 \ea
 \right )
 \,
 \left (
 \begin{array}{c}
 \psi_1^{[1]}\\
 \psi_2^{[1]}\\
 \psi_3^{[1]}\\
 \vdots
 \ea
 \right )
  =
  \ee
 \be
 =
 -\lambda\,
 \left (
 \begin{array}{cccc}
 {\cal V}_{00}&{\cal V}_{01}&{\cal V}_{02}&\ldots\\
 {\cal V}_{10}&{\cal V}_{11}&{\cal V}_{12}&\ddots\\
 {\cal V}_{201}&{\cal V}_{21}&{\cal V}_{22}&\ddots\\
 \vdots&\ddots&\ddots&\ddots\\
 \ea
 \right )\,
 \left (
 \begin{array}{c}
 \psi_0(0)\\
 0\\
 0\\
 \vdots
 \ea
 \right )
 \,.
  \label{spermJBK}
 \ee
In the context of the conventional Rayleigh-Schr\"{o}dinger
perturbation-expansion recipe
this is precisely the equation
which would yield the explicit formula for
coefficient $E^{[1]}$ defined in terms
of the matrix elements of perturbation ${\cal V}$.
Nevertheless, as long as
our present unconventional unperturbed Hamiltonian is
a non-diagonal (and, moreover, infinite-dimensional) matrix,
we have to pay the price: The left eigenvector $\bbr \chi(0)|$
of our unperturbed Hamiltonian is not at our disposal.
We cannot use it for the standard pre-multiplication
of Eq~(\ref{spermJBK}) from the left. This means that
without the knowledge of  $\bbr \chi(0)|$,
the first line of
Eq.~(\ref{spermJBK}), viz., relation
 \be
 E^{[1]}={\cal V}_{00}+\psi_1^{[1]}/\psi_0(0)\,
 \ee
only enables us to
extract the value of $E^{[1]}$ in the form of function of
an unknown
parameter $\psi_1^{[1]}$.
This is the ambiguity which can be perceived as
mimicking the unaccounted influence of
the rest of the matrix elements of perturbation ${\cal V}$.

The
latter formal disadvantage is partially compensated by the
presence of an easily invertible
triangular matrix in
Eq.~(\ref{spermJBK}).
This suggests that
the role of a variable parameter could rather be played, in
a partial resemblance of the EPN recipe,
by the energy correction $E^{[1]}$ itself.
We would then have
$$\psi_1^{[1]}=\psi_1^{[1]}(E^{[1]})=(E^{[1]}-{\cal V}_{00}) \psi_0(0)\,.$$
Similarly,
from the second row of
Eq.~(\ref{spermJBK})
we would 
obtain 
the value of the
second wave-function component
 \be
 \psi_2^{[1]}=(E(0)-E_1)
 \psi_1^{[1]}(E^{[1]})
 -{\cal V}_{10}\psi_0(0)
 \ee
etc.

\section{Discussion\label{ladel}}


We can conclude that in both the EPN- and IEP-related
unitary-evolution scenarios
the properly amended form of perturbation theory
seems to be able to provide,
even in its leading-order
form, some explicit and useful
criteria of the
acceptability or unacceptability of various preselected
perturbations of
phenomenological interest.

\subsection{Benign
perturbations}

Between the EPN and IEP
alternatives
there still exists a crucial difference.
Indeed, in the typical EPN-related analysis
our considerations usually start from
our knowledge of
the ``physical'' family of models $H(g)$.
Then, the only truly
difficult problem is to localize the EPN singularity,
especially when the values of $N$ are not too small.
In the case of the IEP singularities, in
contrast,
we
only know, typically, the unperturbed Hamiltonian
as sampled here by the IC operator.
It is possible to conclude that precisely this
makes the
IEP-related models perceivably
more difficult to study.

From a purely pragmatic point of view
a source of certain optimism
could
be drawn
from
the
leading-order
perturbation-approximation criteria.
Their {}{key} strength
lies in the
possibility of identification of the ``malign''
IEP perturbations which would destroy the reality of the spectrum and
which would make the evolution non-unitary.

The complementary
reliable identification of
the
``benign''
perturbations
{}{is}
a mathematically much more difficult open {}{problem}.
Incidentally, qualitatively the same conclusions have
already been obtained
in the simpler EPN context. For example,
in the
above-mentioned study \cite{UweE} of a
specific Bose-Hubbard model near its EPN dynamical extreme,
the authors did not insist on
the reality of the spectrum.
{}{They decided to treat} their mathematical results as
applicable {}{and valid}
in a broader, not necessarily unitary open-system context.

In a narrower, closed-system setting,
a deeper analysis {}{has been performed}
and a
resolution of the apparent EPN-related
instability paradox {}{has been} described
in paper \cite{corridors}. We studied {}{there} the
exact as well as approximate secular equations in more
detail. Our ultimate conclusion was that the necessary smallness
condition specifying the class of the admissible, unitarity
non-violating perturbations does not involve their upper-triangular
matrix part at all.
In contrast, for the perturbed-EPN
model in question, the lower-triangular matrix part
of all of the ``benign''
(i.e., unitarity-compatible) perturbations
has been shown to have
the following element-dependent matrix form of
condition of the sufficient smallness {}{of
 $\lambda$},
 \be
 \hspace{-1.6cm}\lambda\,V^{(N)}_{(admissible)}=\left[ \begin {array}{cccccc}
  {\lambda}^{1/2}{{\mu}}_{11}
  &0&\ldots&0&0&0
  \\\noalign{\medskip}\lambda\,{\mu}_{{21}}&{\lambda}^{1/2}{{\mu}}_{22}
&\ldots&0&0&0
  \\\noalign{\medskip}{\lambda}^{3/2}
  \,{\mu}_{{31}}&\lambda\,{\mu}_{{32}}&\ddots&\vdots&\vdots&0
  \\\noalign{\medskip}{\lambda}^{2}{\mu}_{{41}}&{\lambda}^{3/2}
  \,{\mu}_{{42}}
  &\ddots&\ddots&0&0
  \\\noalign{\medskip}\vdots&\vdots&\ddots&\lambda\,{\mu}_{{{N}-1{N}-2}}&
  {\lambda}^{1/2}{{\mu}}_{{N}-1{N}-1}
 &0
  \\\noalign{\medskip}{\lambda}^{{N}/2}{{\mu}}_{{N}1}&
 {\lambda}^{({N}-1)/2}{\mu}_{{{N}2}}&\ldots&\lambda^{3/2}
 \,{\mu}_{{{N}{N}-2}}&\lambda\,{\mu}_{{{N}{N}-1}}&{\lambda}^{1/2}{{\mu}}_{{N}{N}}
 \end {array} \right]\,.
 \label{uho}
 \ee
The matrix structure
(\ref{uho}) may be interpreted as manifesting a characteristic
anisotropy and the hierarchically ordered weights of influence of
the separate matrix elements
because during the decrease of $\lambda\to 0$, all of the ``benign''
matrix-element parameters have to have bounded components
$\mu_{j,k}={\cal O}(1)$.
For a more explicit explanation we may rescale
 \be
 \lambda\,V^{(N)}_{(admissible)}=\lambda^{1/2}\,B(\lambda)\,
 V^{(reduced)}\,B^{-1}(\lambda)
 \label{216}
 \ee
where $B(\lambda)$ would be a diagonal matrix with elements
$B_{jj}(\lambda)=\lambda^{j/2}$ and where the whole
reduced ``benign'' matrix of perturbation
would be bounded, $V^{(reduced)}_{jk}={\cal O}(1)$.

On this necessary-condition background valid at all dimensions $N$,
the samples of sufficient conditions retain a purely numerical
trial-and-error character, with the small$-N$ non-numerical
exceptions discussed, in \cite{corridors}, for the matrix dimensions
up to $N=5$.

\subsection{IC oscillator as popular toy model}

In order to elucidate the
benchmark-model
role of the IC IEP oscillator
let us recall
paper \cite{BB} in which
Bender with Boettcher
examined a
rather broad family of time-independent non-Hermitian
toy-model Hamiltonians (cf. Eq.~(\ref{bebino}) in Appendix A. 2 below).
They felt guided by the
postulate of (antilinear) symmetry of their models,
 \be
  {\cal PT}H^{(BB)}=H^{(BB)} {\cal PT}\,.
 \label{psh}
 \ee
The linear operator ${\cal P}$
was treated as parity (causing
the space reflection $x \to -x$)
while ${\cal T}$ {}{had} to mimic the
anti-linear time reversal.

{}{The} authors
proposed to treat their operators $H^{(BB)}$ as ``Hamiltonians
whose spectra are real and positive''
so that ``these ${\cal PT}-$symmetric theories
may be viewed as analytic continuations of conventional
theories from real to complex phase space''
\cite{BB}.
During the subsequent
wave of development of the related mathematics
it has been revealed
that
in the language of functional analysis
the
${\cal PT}-$symmetry {}{of Eq.~(\ref{psh})}
can be re-read as
pseudo-Hermiticity
\cite{ali}
as well as a
self-adjointness in the Krein space {}{endowed}
with indefinite pseudo-metric ${\cal P}$
\cite{Langer,book,Joshua}.

A deeper mathematical insight in the
class of ${\cal PT}-$symmetric
models has been obtained.
In the narrower context of quantum mechanics
of closed systems, in contrast,
the IEP-possessing IC model itself has not been assigned, up to now, any
sufficiently
consistent phenomenological interpretation yet \cite{UweS}.
Still, in retrospective,
its temporary popularity
was enormous. Its roots may be dated back
to the Bessis' and Zinn-Justin's
unpublished \cite{DB}
but
widely communicated \cite{BB}
discovery
that in spite of the manifest non-Hermiticity of the IC Hamiltonian
its spectrum appeared to be real and bound-state-like, i.e.,
discrete and bounded from below.

In the extensive existing literature devoted to the study
of systems with ${\cal PT}$ symmetry (cf., e.g., reviews
\cite{Carlbook,Christodoulides}), a lot of attention has been paid to
the
non-Hermitian but still sufficiently realistic
ordinary differential Hamiltonians of the form $H = T + V$
reminiscent of the IC oscillator
in which
the entirely conventional kinetic-energy term $T=-d^2/dx^2$
is combined with a suitable local complex one-dimensional
potential $V = V(x)$.
By many authors
the latter models were sampled by the
field-theory-mimicking
oscillator Hamiltonian (\ref{SE})
in which the purely imaginary form of the
asymptotically growing potential is a
truly puzzling mathematical curiosity.

This was also the feature which attracted a lot of attention
among physicists \cite{BB,Carl,book,Carlbook}.
Precisely because they happened to
generate the purely real, discrete and
non-negative (i.e., hypothetically, observable and
bound-state-like)
energy spectra.
Still, the ultimate verdict
by mathematicians \cite{Siegl,UweS}
was discouraging
because
they proved
that
the IC Hamiltonian
cannot be assigned any
isospectral
self-adjoint avatar $\mathfrak{h}({{t}})$
or acceptable physical
inner-product metric \cite{Siegl}.
Thus, the rigorous mathematical
analysis finally led to
the
loss of some of the most optimistic phenomenological
expectations.

\section{Summary\label{bladel}}

Not too surprisingly, the highly desirable proofs of the so called
unbroken form of ${\cal PT}-$symmetry
(in which, by definition \cite{Carl}, the spectrum remains real)
appeared to be, in numerous
applications, strongly model-dependent.
There seemed to be no
universal criteria guaranteeing
the existence
of the
unbroken ${\cal PT}-$symmetry
in
dependence
on
a suitable measure {}{of} degree of
the
non-Hermiticity of the Hamiltonian.

During the preparation of our present study
we came to the conclusion that
the lack of
a deeper
understanding of correspondence
between the (apparent) non-Hermiticity and
(hidden) unitarity
might have been caused by
an overambitious generality of the choice of the
models in the literature.
For this reason we decided to
narrow the scope of our analysis
and we decided to restrict our attention
just to the extremely non-Hermitian
Hamiltonians
which would lie very close to their EPN or IEP singularity.

In our present paper we explained that and how
such a decision enabled us
not only to pick up a
rather natural measure of
the non-Hermiticity
(characterized simply by the inverse distance of
{}{the variable physical parameter} $g \in {\cal D}$
from its {}{unphysical} exceptional value)
but also
to formulate a
well-defined project
in which we developed and applied, consequently,
some innovative and suitable perturbation-approximation techniques.
In its framework we managed to show that
the
unbroken ${\cal PT}-$symmetry
of our models can really survive
inside an open parametric domain,
on a point of boundary of which
our measure of
non-Hermiticity
reaches its maximum.
The latter point
{}{(which remains
manifestly unphysical)}
has been shown to coincide either with the
Kato's \cite{Kato} exceptional point of
a finite order $N$
or with its hypothetical IEP analogue.

Such an approach has been found productive.
Using certain slightly modified
techniques of
perturbation theory
of linear operators {}{with finite $N$}
we found that, paradoxically,
the restriction of attention
to the smallest
vicinity of the singularity
(in which the Hamiltonians
become maximally non-Hermitian)
leads to a remarkable simplification
of the perturbation-approximation constructions.
{}{In spite of being singular and
unacceptable
as observables at $g=g^{EPN}$},
the special, ``exceptional''
non-diagonalizable operators
appeared to be
eligible
as unperturbed Hamiltonians.
In their
vicinity  {}{such that $g \in {\cal D}$}
their diagonalizability as well as
the
observability
status {}{(i.e., the standard
physical
status)}
were
re-established.

{}{In this sense, the core of our present message is that
the same perturbation-regularization physical interpretation
should be also attributed to the IEP models where $N=\infty$}.
{}{For the purposes of illustrative example we choose
the popular imaginary oscillator Hamiltonian.
Such a choice has been found motivated, first of all, by its
long-lasting theoretical significance
which ranges from
its more or less purely formal role in mathematics
and functional analysis
\cite{usulum}
up to a deeper
phenomenological significance in quantum statistics
(where the imaginary $\phi^3$ interaction
mimics the Yang-Lee edge singularity
\cite{BB,Fisher}) and
up to its important theoretical role
of a benchmark model
in the conformal quantum field theory
\cite{usulumb}
as well as in the less well known but still popular Reggeon
field theory
\cite{usulumc}}.

{}{In all of these contexts
our present results imply that the IEP property of
the IC-like models means
unphysicality.
Only a suitable perturbation can reinstall the (possibly, ``hidden''
{\it alias\,} ``quasi-'')
unitarity and physicality.
Thus, the
practical realizations of the standard quantum-mechanical
IC model
remain, in a way and for the reasons as outlined in \cite{Siegl},
elusive, in the unitary-theory context at least \cite{Nimrod}.
At the same time
one might still expect that some of its
realizations could emerge off the realm of
quantum mechanics, i.e., say, in optics \cite{Christodoulides}.
}

Constructively we managed to
clarify also some of the consequences
{}{of our present
perturbative-regularization recipe}.
The emergence of
qualitative as well as quantitative EPN - IEP parallels
helped us to complement and understand better
the
twelve years old disproof of the
internal mathematical consistency of
the IC IEP quantum oscillator \cite{Siegl}.
Such a clarification
can be perceived as being of a
fundamental importance in quantum theory.
Indeed, potentially, most of our observations
might immediately be extended
also to many other currently
popular {}{but IEP-singular} non-Hermitian quantum
models.

\newpage

\section*{Appendices}

\subsection*{A. 1. Paradox of stable bound states in complex potentials}

For a long time it was believed that
the locality of the real and confining potential is so
strongly restrictive a constraint
that the loss of the reality of $V(x)$
(i.e., of the self-adjointness of
the Hamiltonian in any standard Hilbert space
of states, i.e., say, in $L^2(\mathbb{R})$)
would immediately imply the loss of the reality
of the spectrum,
i.e.,
the loss of the observability of the quantum system
in question.

In 1998, in the
Bender's and Boettcher's
pioneering letter \cite{BB}
the latter belief has been
strongly opposed.
Using a combination of methods
these authors
argued that also the spectrum generated by
multiple {\em manifestly complex\,} local
interaction potentials $V(x)$
still appears to be strictly real and discrete,
i.e.,
fully compatible
with the conventional postulates of quantum mechanics of
the stable and unitary bound-state quantum systems.

Subsequently, the proposed amendment
of the model-building
paradigm has widely been accepted.
For various complex $V(x)$s,
rigorous \cite{DDT,Shin}
as well as numerical \cite{Rafa}
proofs of the reality of the spectra
were found
and attributed to a certain ``hidden form of Hermiticity''
of the underlying Hamiltonians (cf., e.g., a few earlier
review papers \cite{ali,Carl} for details).

A return to older literature
(cf., e.g., review \cite{Geyer}) revealed that
the compatibility of the unitarity of the
evolution with a manifest non-Hermiticity
of the interaction
can be given a comparatively elementary explanation
because whenever the Hamiltonian $H$ in question
has a real and discrete spectrum,
it may be {safely self-adjoint\,}
with respect to another,
``correct'', {\it ad hoc\,} inner product, i.e., in a
modified, ``physical''
Hilbert space ${\cal H}_{phys}$.
{Simultaneously}, it may make sense to
stay working in the initial and more user-friendly
Hilbert space ${\cal H}_{math}$ which remains
``unphysical''
(i.e., formally non-equivalent)
but, for some reasons, preferred.

Many years ago
many people really
studied various toy models
of such a type, characterized by
the interaction which appeared {\em manifestly non-Hermitian\,}
with respect to a conventional
inner product.
The scope of such -- mostly, numerical -- attempts
ranged from very pragmatic Dyson-inspired analyses of
non-relativistic
many-body systems \cite{Dyson,Jenssen}
up to the abstract, methodically motivated
considerations concerning the applicability
of non-Hermitian models in the
relativistic quantum field theory
\cite{DB}.

In the latter context,
the Bender's and Boettcher's results \cite{BB}
proved particularly inspiring and made the idea popular.
In parallel,
the most elementary IC model appeared to represent
a challenge in mathematics,
leading, i.a., to
a rigorous proof of
the reality of its spectrum
by Dorey et al \cite{DDT}.
For this reason
the model served, for many years,
as a benchmark
methodical guide which inspired
several new developments
in relativistic quantum field
theory \cite{BM}
as well as in multiple other
phenomenologically oriented subdomains of
modern physics \cite{Christodoulides,Nimrod}.

Last but not least,
Bender with Boettcher
extended the spectrum-reality
conjecture to a broad class
of potentials
$V^{(BB)}(x)=({\rm i}x)^\delta\,x^2$
with arbitrary
non-negative $\delta \in (0,\infty)$ and
with $x$ lying on a complex contour \cite{BB}.
All of these results caused
the growth of the popularity of the
innovative reformulation of quantum physics
of unitary systems
admitting manifestly non-Hermitian Hamiltonians
among physicists.
This, not too surprisingly,
appeared paralleled
by a
criticism
by mathematicians who
referred, e.g., to the
existence of counterexamples
with pathological properties \cite{Trefethen}. Incidentally,
some of these counterexamples were even already known,
many years earlier, to
Dieudonn\'{e} \cite{Dieudonne}).

Fortunately,
{}{an ultimate}
resolution of the conflict
has  been found in a rediscovery and return to
a half-forgotten
but still fully relevant older review
paper by Scholtz et al \cite{Geyer}.
In it, most of the objections by mathematicians
were circumvented by an {\it ad hoc\,} technical assumption
that one is only allowed to consider the non-Hermitian Hamiltonians
(as well as any other candidates for observables)
which are, as operators in Hilbert space, bounded:
cf. also several mathematically oriented
reviews in \cite{book} in this respect. Thus, one may
call such a mathematically consistent
version of the theory quasi-Hermitian quantum mechanics.

\subsection*{A. 2. Beyond the imaginary cubic-oscillator potential}

The
above-mentioned Bender's and Boettcher's
choice of the illustrative stationary
non-Hermitian one-dimensional
(i.e., ordinary differential)
Hamiltonians
 \be
 H^{(BB)}_{}=-\frac{d^2}{dx^2} +V_{(\delta)}(x)\,,
 \ \ \ \
 V_{(\delta)}(x)=
 ({\rm i}x)^\delta\,x^2\,,
 \ \ \ \
  \delta \in (0,\infty)\,
  \label{bebino}
 \ee
has been motivated by
their conjecture that
the ubiquitous requirement of the
self-adjointness of the observables
might be criticised
as over-restrictive.
They proposed that one should
consider a broader class of
Hamiltonians
$H$
for which the conventional
condition
of
self-adjointness  becomes replaced
by the
property
called
${\cal PT}-$symmetry (cf. Eq.~(\ref{psh}))
{\it alias\,} ${\cal P}-$pseudo-Hermiticity,
 \be
  {\cal P}H^{(BB)}=\left [H^{(BB)}
  \right]^\dagger {\cal P}\,.
 \label{upsh}
 \ee
Under the latter assumption (cf. also
the comments in \cite{Langer,BG})
Bender with Boettcher
assumed that the role of the
guarantee of the reality of the spectrum
of the bound-state energies
(i.e., in principle, of their observability)
can be relegated from the conventional Hermiticity
to
the ${\cal PT}-$symmetry of the system
whenever such a symmetry
remains spontaneously unbroken \cite{Carl}.

In applications
the choice of ${\cal PT}-$symmetric
Hamiltonians
appeared strongly influenced by
a tacit reference to the principle of correspondence
due to which  $H$ is assumed split into its
kinetic-energy component  $H_{kin}$
and a suitable interaction term $H_{int}$.
Moreover,
the analysis is often
restricted just
to the
single-particle one-dimensional motion
with conventional $H_{kin}\sim -d^2/dx^2$
and with a suitable
local-interaction form of $H_{int}\sim V(x)$.

This choice
has already been recommended by Bender
with multiple collaborators (cf. review \cite{Carl}).
They
emphasized
that the study of various non-Hermitian
but ${\cal PT}-$symmetric
quantum models
with real spectra
can be perceived as
motivated by
quantum field theory.
In this context a key role is played,
in a way proposed by
Bessis with collaborators \cite{DB},
by the imaginary cubic (IC) potential
$V^{(IC)}(x)={\rm i}x^3$.
For this reason,
Siegl with Krej\v{c}i\v{r}\'{\i}k \cite{Siegl}
turned their attention to the IC
Hamiltonian (\ref{SE}), having
revealed that
such a model suffers of unpleasant
pathologies.

These pathologies appeared
only too serious to be ignored.
After all,
Siegl with Krej\v{c}i\v{r}\'{\i}k only
rediscovered the Dieudon\'{e}'s older claim
that for such a Hamiltonian
``there is for instance
no hope of building functional calculus that would follow
more or less the same pattern as the functional calculus
of self-adjoint operators''  \cite{Dieudonne}.
Siegl with Krej\v{c}i\v{r}\'{\i}k also
listed
several ``pathological properties
of non-self-adjoint  $H^{(IC)}$'' and
they offered a
rigorous proof that these features of the IC model
find a close formal
analogue
in
the Kato's EPNs.

Thus,
the popular toy-model
operator $H^{(IC)}$
became
disqualified
as a candidate
for quantum Hamiltonian (see also an independent reconfirmation
of the skepticism, say, in \cite{UweE,Semoradova} and,
after all, also in the very first line
of the abstract of the comprehensive review \cite{ali}
requiring the diagonalizability of the observables).
Still, several attempts were made to
replace $H^{(IC)}$ by a suitable regularized alternative.
Typically, the regularization
has been sought in
a truncation of the real line of $x$
(cf. \cite{Tretter}).
Unfortunately,
one can hardly speak about a successful resolution
of the problem
because
one form of unacceptability
was merely replaced by another one,
viz., by the complexification of the spectrum.

\subsection*{A. 3. Beyond the stationary quasi-Hermitian models}

What is most characteristic for
the
applications of quantum mechanics
in the so called Schr\"{o}dinger picture \cite{Messiah}
is the observability of the
generator of the evolution
of wave functions 
called ``Hamiltonian''.
In most applications
it is required self-adjoint in the preselected
Hilbert space ${\cal H}$.
Then, according to Bender and Boettcher \cite{BB},
the robust nature of the
reality of
its eigenvalues (representing, in many models,
just the discrete bound-state energies)
can be perceived as
a weakness of
the approach. Indeed,
once we prepare,
at an initial time $t=0$, the system
in a pure state ({\it alias\,} ``phase'')
described by a ket-vector $|\psi(0)\kt \in  {\cal H}$,
we discover that the ``phase''
(defined by the specific set of observable aspects)
cannot be changed
by the evolution.

This feature would make the description of
quantum phase transitions impossible.
Fortunately, a change of the ``phase''
(i.e., e.g., an abrupt loss of the observability
of the time-dependent bound-state energies)
has recently been rendered possible after
a conceptually straightforward
transition from the Schr\"{o}dinger-picture (SP) approach
to a formally equivalent,
albeit technically more complicated
non-Hermitian interaction picture
(NIP, \cite{timedep}). 

Some of the consequences of such a
change of paradigm
become relevant also for an appropriate
understanding and treatment of
the IEP-related considerations.
Due to the lack of space for an exhaustive analysis
of this problem, let us only briefly mention that
in the traditional and most popular SP framework
of conventional textbooks
the unitary
evolution of a closed quantum system
is just being described
in
a unique preselected Hilbert space ${\cal H}$.
People also often accept multiple additional {\it ad hoc\,}
simplifying assumptions,
with the most popular one
concerning the above-mentioned generator of evolution
of wave functions (say, $G_{(textbook)}(t)=\mathfrak{h}(t)$)
and requiring its self-adjointness in ${\cal H}$,
 \be
 \mathfrak{h}(t) = \mathfrak{h}^\dagger(t)\,.
 \label{herm}
 \ee
Another such a traditional simplification
concerns
the inner product in ${\cal H}$
which is assumed
time-independent \cite{Geyer,ali}.

In the generalized NIP framework, in contrast,
one has to consider
Schr\"{o}dinger equation
 \be
 {\rm i}\frac{d}{dt}\,|\psi(t)\kt = G(t)\,|\psi(t)\kt\,,
 \ \ \ \ \ |\psi(t)\kt \in {\cal H}\,
 \label{tdse}
 \ee
in which the generator $G(t)$
need not represent an observable
\cite{Fring,NIP,Maamache,Bishop,Mousse}.
From the
purely phenomenological point of view
such a generalization is useful.
In a way reflecting the widespread
knowledge of the above-sampled differential-operator benchmark
models (\ref{bebino}) it is still possible
to introduce the energy-representing observables
 \be
 H(t) = \triangle + V(x,t)
 \label{super}
 \ee
which are not only
non-Hermitian and
manifestly time-dependent
but also
different from the generator $G(t)$.
In this setting
the freedom of choice between
the SP or NIP framework only means that
one treats the time-dependence of our observables
as
inessential or essential, respectively.

In practice we usually insist on the
standard
phenomenological and probabilistic interpretation
and, in particular, on the
observable energy status of the
specific operator (\ref{super}).
Thus, we have to keep in mind
that at least
some of the most popular differential
operators
cannot be used as benchmark models without hesitation.
For this reason,
a consequent constructive realization of
description of
the phenomenon of
a genuine quantum
phase transition
remains to be a task
for the future development of the  theory.

\subsection*{A. 4. The question of the unitary-evolution accessibility of EPNs}

Due to the degeneracy
of the unperturbed energy spectrum in
EPN limit the $N-$plet of the perturbed-energy roots
of the corresponding secular equation
(cf., e.g., the
bound-state
energy roots
$\epsilon_n
=\epsilon_n(\lambda)$ of Eq.~(\ref{krutadef})
with $n=1,2,\ldots,N$)
need not necessarily be all real
and, hence,
representing observable quantities.
In Refs.~\cite{UweE,admissible},
for example, even some of the approximate
leading-order roots were
found complex.
This observation can be reinterpreted as indicating that
even in an immediate
EPN vicinity even the bounded perturbations may
still be reclassified, in unitary theory, as ``inadmissibly large'',
forcing the system to perform an abrupt
quantum phase transition.

Within quantum mechanics
of unitary, closed systems
in its
quasi-Hermitian formulation a key to
the suppression of such a quantum catastrophe
lies in
the construction
of a correct physical
inner product in Hilbert space \cite{Geyer}.
Still, many of
the truly realistic applications
of the quasi-Hermitian operators
may remain, in the model-building context, counterintuitive.
In particular, the doubts emerge in
virtually all of the tentative quasi-Hermitian
descriptions of the
phenomenon of quantum
phase transition because,
traditionally, all of such processes have been treated
as non-unitary,
requiring
an {\it ad hoc\,} effective-operator
approach \cite{Feshbach}.

A feasible way out of the apparent deadlock
is offered by the quasi-Hermitian
quantum models
in which a given observable
with real spectrum (say,
$\Lambda(t)$)
is non-Hermitian.
In such a
theory
(cf., e.g., its reviews \cite{Geyer,ali,SIGMA,Carl,book})
the condition of self-adjointness
of $\Lambda(t)$s survives ``in disguise'', being
replaced by a formally equivalent
quasi-Hermiticity condition in another Hilbert space,
 \be
 \Lambda^\dagger(t)\,\Theta(t)=\Theta(t)\,\Lambda(t)\,.
 \label{dieut}
 \ee
The
assumption of the time-dependence of the
related inner-product-metric $\Theta(t)$
opens then the possibility
of reaching a singularity via unitary evolution.

In such a case the
collapse is simply
rendered possible
 by the coherent, simultaneous loss of the existence of
the time-dependent
inter-twiner $\Theta(t)$ in the critical limit of $t \to t^{(EPN)}$
or $t \to t^{(IEP)}$.
One could also say  that
the realization of the whole process of the
change of phase, i.e., of
the loss of the observability
of some of the measurable characteristics
(i.e., of the loss of the quasi-Hermiticity of $\Lambda(t)$)
is to be mediated by
the metric $\Theta(t)$ in (\ref{dieut}) which becomes,
in the limit,
non-invertible and, in fact, just a rank-one operator \cite{passage}.

The emergence of a fully explicit
conflict between the constructive
feasibility and mathematical consistency
can be traced back to the year 2012 and paper \cite{Siegl}
in which
Siegl with Krej\v{c}i\v{r}\'{\i}k
disproved
the acceptability
of
a broad class of the currently popular
non-Hermitian but observable
Hamiltonians.
The impact of the disproof
was truly destructive. The currently
widespread belief
in the benchmark role
of many
non-Hermitian but still observable
Hamiltonians
with real spectra
has been shattered.

In parallel,
the doubts were also thrown upon the
acceptability
of the specific benchmark ordinary-differential
(i.e., one-dimensional and
mathematically still sufficiently user-friendly)
non-Hermitian
candidates for the energy-representing
Hamiltonians
decomposed
into their two intuitively plausible
(and stationary as well as non-stationary)
kinetic- plus potential-energy
components,
 \be
 H=-\frac{\hbar}{2m(x)}\frac{d^2}{dx^2}+V(x)\neq H^\dagger\,.
 \label{assam}
 \ee
The no-go theorems of paper \cite{Siegl}
(see also \cite{Viola} for further details)
seemed to return us back to the
older methodical
analyses
in which
the deepest source of the mathematical
difficulties has been attributed to
the unbounded-operator nature of the most popular
differential operators as sampled by Eq.~(\ref{assam})
(cf., e.g., \cite{Dieudonne,Geyer,Kretschb}).

In applications, paradoxically, the latter disproofs
and skepticism motivated a rapid increase of interest
in the study of the so called open quantum systems
\cite{Nimrod}.
In such a very traditional context
an enormous acceleration of the progress
(say, in an innovated understanding of
the dynamics of resonances)
has been achieved
due to the
successful
applications of the
new methods of the solution of the
non-Hermitian versions of
the Schr\"{o}dinger-like evolution equations.
At present,
multiple branches
of physics
were enriched by these tendencies, including even the
non-quantum ones
\cite{Carlbook,Christodoulides}.

\subsection*{A. 5. A note on the broader quantum-physics framework}

Even in the context of quantum physics,
paradoxically, the
intensification of interest
in non-Hermitian Hamiltonians of closed systems
accelerated the progress
in the development of the
dedicated mathematical methods
and, in particular, in the understanding of non-Hermitian
operators with the spectrum which was not real.
Paradoxically, these developments redirected
attention of a part of physics community
back to the traditional models in which the
meaningful
spectra were allowed
to be complex.

Our lasting interest in
unitary
quantum mechanics using hiddenly Hermitian observables
is partly motivated,
in a way
documented in paper \cite{catast},
by the possibility of
mimicking the processes of quantum phase transitions.
We believe that such a direction of analysis
must necessary profit from the very recently introduced
combinations of the requirements of the manifest
non-Hermiticity (and, especially, of its hiddenly Hermitian
forms called quasi-Hermiticity \cite{Dieudonne,Geyer})
with some other suitable
model-building options
and auxiliary technical assumptions like the time-dependence of the
operators and/or
their
${\cal PT}-$symmetry
(cf. \cite{BB,Carl,BG}) or factorization (cf., e.g., \cite{BG,hybrid,[380]}).

The main stream of
our considerations remained restricted to the
context of
quantum physics
and quasi-Hermitian dynamics
in which
we insisted on
the compatibility
of our models
with all of the basic principles of quantum
mechanics of the so called closed and unitary systems
admitting the standard probabilistic
interpretation.
This does not mean that
the scope of the theory
and of its applications cannot be much broader,
in principle at least.
Innovations may be obtained in the representation
of the states as well as of their observable characteristics.

In the spirit of multiple relevant recent reviews
this goal can be achieved by the
various physical reinterpretations
of the parameter-dependent Hamiltonians $H(g)$.
Even when we only admit,
in
Schr\"{o}dinger picture, its observable-energy
interpretation,
it is still worth returning to the
Dyson's treatment \cite{Dyson} of such an operator
as the one which is isospectral
with its self-adjoint avatar $\mathfrak{h}({{g}})$,
 \be
 H({{g}}) \to \mathfrak{h}({{g}})=
 \Omega({{g}})\,H({{g}})\,\Omega^{-1}({{g}})
 =\mathfrak{h}^\dagger({{g}})\,.
 \label{redun}
 \ee
In this manner, even the metric $\Theta$ itself acquires
an entirely new meaning of the mere product
  \be
 \Theta(g)=\Omega^\dagger{({g})}\,
 \Omega({{g}})
 \label{aspro}
 \ee
of the so called Dyson's maps which are
non-unitary and
related to the
conventional quantum physics avatar $\mathfrak{h}({{g}})$
rather than to the non-Hermitian
upper-case Hamiltonian itself.

In this spirit, Dyson introduced
and treated the mappings $\Omega$ in (\ref{redun})
as certain variationally motivated {\it ad hoc\,}
multiparticle correlations.
In contrast, Buslaev with Grecchi \cite{BG}
offered and formulated another point of view by which
these operators represent just
an isospectrality-equivalence transition from a Hilbert space
which is unphysical to
another Hilbert space
which is physical.
During such a transition
it is possible to distinguish and cover both the
open quantum systems (in which one describes resonances)
and the closed quantum systems
(in this case, in {\it loc. cit.}, Buslaev with Grecchi
paid their attention to the quartic anharmonic oscillators).

In the latter, newer and less traditional case
one has to
construct the auxiliary metric
as product (\ref{aspro}).
This just
amends the inner product in
the mathematically friendlier and  computationally preferred
but manifestly
unphysical Hilbert space ${\cal K}$,
 \be
 \br \psi_1|\psi_2\kt_{in\ physical\ space}
 =\br \psi_1|\Theta\,|\psi_2\kt_{in\ mathematical\ space}\,.
 \label{relat}
 \ee
In this notation the obligatory condition of the
Hermiticity of $\mathfrak{h}({{g}})$
(cf. Eq.~(\ref{redun}))
becomes translated into the equally obligatory condition of the
quasi-Hermiticity of $H({{g}})$ in the mathematical Hilbert space,
 \be
 H^\dagger({{g}})\,\Theta({{g}})=\Theta({{g}})\,H({{g}})\,.
 \label{quasihe}
 \ee
Thus, for a preselected Hamiltonian $H(g)$
with real spectrum,
its acceptability as a closed-system observable
will be guaranteed either by the
Hermiticity of its
$\Omega-$transformed isospectral
avatar
or, equivalently,
by the $\Theta-$quasi-Hermiticity property
of $H(g)$ itself.

\subsection*{A. 6. Final note on the notation and outlook}

In the literature
devoted to the models using quasi-Hermitian
observables the notation conventions
did not unite yet. Thus, the Hilbert-space metric
(which we decided to denote by
the upper-case Greek symbol $\Theta$)
can be found denoted
as $T$ (which does not mean time reversal,
cf. equation Nr. (2.2)
in one of the oldest reviews \cite{Geyer})
or as subscripted lower-case Greek $\eta_+$ (cf. equation Nr. (52)
in one of the more modern reviews \cite{ali})
or as $\exp(-Q)$ (cf. the 2006 paper \cite{Mateo})
or as $\rho$ (cf. \cite{Frith})
or by the letter $G$ (cf. Tables Nr. I and II in \cite{Ju}), etc.

The notion of EPs (exceptional points)
of our present interest
emerged
within the strictly mathematical
theory of linear operators \cite{Kato}.
It
played there a
key role in the rigorous analysis
of the criteria of convergence of perturbation series.
In the context of physics,
the notion
was less well known, being called there the
Bender-Wu singularity \cite{Alvarez} etc.
Independently,
this notion has only been found important
for physicists \cite{Heiss,Heissb},
especially during
the last twenty years, viz., during the
growth of interest in the role played by the
non-self-adjoint operators in several
(i.e., not always just quantum) branches
of phenomenology
\cite{Carlbook,Christodoulides}.

During the early stages of development of the
latter innovative approach
the role of a benchmark illustrative example
has been played
by the IC (imaginary cubic)
differential-operator Hamiltonian
of
Eq.~(\ref{SE}).
Later,
as we already explained above,
such a choice of illustration
proved to be a bit unfortunate.
For proof we cited
Siegl and Krej\v{c}i\v{r}\'{\i}k,
\cite{Siegl}
who emphasized that, in the formal sense, the
obstacles imposed by the loss
of the Riesz-basis diagonalizability
of the IC Hamiltonian are
``much stronger than'' those imposed by ''any
exceptional point
associated with finite Jordan blocks''.

The related
concept of asymptotic IEP
was not only very new but also rather elusive. Even its
definition as
provided by the authors was
just implicit
(see
sections
IID, IIIC and IV in \cite{Siegl}).
An explanation is that
their
message
was aimed, first of all, at the community
of physicists for which the IEP IC oscillator model
served as a
heuristic ``fons and origo'' of what
has been widely accepted as ${\cal
PT}-$symmetric quantum mechanics.
The same
authors also
emphasized that the
properties of $H^{(IC)}$
``are essentially different with respect to self-adjoint
Hamiltonians, for instance, due to spectral instabilities''.
Thus,
the main
IEP-related result of \cite{Siegl} (viz., the proof of
the existence of an IEP anomaly in the IC model)
was finally formulated
as an observation that
``there is no quantum-mechanical
Hamiltonian associated with it
via \ldots similarity transformation''.

The latter conclusion was revolutionary
and opened a number of
new questions concerning the necessity of finding ``new directions
in physical interpretation'' of the model.
In our present paper we, perhaps, threw new
light on the issue, with a rather sceptical conclusion
that
the currently unresolved status of
the twelve years old conceptual task of the interpretation
of the IEP-related
instabilities
does not seem to have an easy resolution, indeed.

\end{document}